\documentclass{article}

 \usepackage[dblblindworkshop, final]{neurips_2025}
\workshoptitle{AI for Music}

\usepackage[utf8]{inputenc} 
\usepackage[T1]{fontenc}    
\usepackage{hyperref}       
\usepackage{url}            
\usepackage{booktabs}       
\usepackage{amsfonts}       
\usepackage{nicefrac}       
\usepackage{microtype}      
\usepackage{xcolor}         
\usepackage{graphicx}
\usepackage{booktabs,tabularx}
\usepackage{amsmath}
\usepackage{booktabs,threeparttable}
\usepackage{subcaption}

\usepackage{hyperref}
\usepackage{url}
\usepackage{graphicx}

\usepackage{inconsolata}
\usepackage{array}
\usepackage{pifont}
\usepackage{tabularx}
\usepackage{adjustbox}
\usepackage{multirow}
\usepackage{enumitem}
\usepackage{xspace}
\usepackage{tcolorbox}
\usepackage{booktabs,subcaption,amsfonts,dcolumn}
\usepackage{url}
\usepackage{amsmath,amsthm,amsfonts,amssymb,bm,stmaryrd, bbm}
\usepackage{color,xcolor,colortbl}
\usepackage{CJKutf8}
\usepackage{makecell}
\usepackage{booktabs}
\usepackage{graphicx}
\usepackage{listings}
\usepackage{float}
\usepackage{wrapfig}
\usepackage{booktabs,siunitx,threeparttable}
\usepackage{xcolor}
\definecolor{Mycolor-green}{HTML}{CDE8CD}
\definecolor{Mycolor2}{HTML}{DDEEFA}
\definecolor{Mycolor-red}{HTML}{FCEAEA} 

\title{Composer Vector: Style-steering Symbolic \\Music Generation in a Latent Space}

\author{%
  Xunyi Jiang, Mingyang Yao, Jingyue Huang, Julian McAuley\thanks{Corresponding Author.} \\
  Department of Computer Science and Engineering\\
  University of California, San Diego\\
  La Jolla, CA 92092 \\
  \texttt{\{xuj003, m5yao, jih150, jmcauley\}@ucsd.edu} \\
}

\begin{document}

\maketitle

\begin{abstract}
Symbolic music generation has made significant progress, yet achieving fine-grained and flexible control over composer style remains challenging. Existing training-based methods for composer style conditioning depend on large labeled datasets.
Besides, these methods typically support only single-composer generation at a time, limiting their applicability to more creative or blended scenarios. 
In this work, we propose \textbf{Composer Vector}, an inference-time steering method that operates directly in the model’s latent space to control composer style without retraining.
Through experiments on multiple symbolic music generation models, we show that Composer Vector effectively guides generations toward target composer styles, enabling smooth and interpretable control through a continuous steering coefficient.
It also enables seamless fusion of multiple styles within a unified latent-space framework.
Overall, our work demonstrates that simple latent-space steering provides a practical and general mechanism for controllable symbolic music generation, enabling more flexible and interactive creative workflows. \footnote{Codes are available in \url{https://github.com/JiangXunyi/Composer-Vector}.}
\footnote{Demo page: \url{https://jiangxunyi.github.io/composervector.github.io/}}

\end{abstract}

\section{Introduction} 
Symbolic music generation can now produce convincing melody, harmonies, and large-scale structures \citep{lu2023musecocogeneratingsymbolicmusic, bhandari2024text2midigeneratingsymbolicmusic, wangNotaGenAdvancingMusicality2025b}.
Despite recent advances, controllable and fine-grained generation remains challenging \citep{tian2025xmusic}. 
Among various control dimensions, composer-style generation stands out as a particularly difficult case \citep{DBLP:journals/corr/abs-2506-17497}.
Unlike music attributes such as tempo, key signature, and velocity, a classical composer's style is often data-scarce since each composer typically has only a limited number of available pieces. 
Apart from data, the style of a composer contains a high-level combination of multiple musical factors, including melody, harmony, texture, and rhythm to make them unique \citep{10734713}.
This combination of data sparsity and representational complexity makes it difficult for traditional supervised or conditioning-based methods to stably control composer style while maintaining model generality.
Existing approaches for composer-controlled generation (focusing on classical music in this work) predominantly rely on training-based methods \citep{10734713, DBLP:journals/corr/abs-2506-17497}, which is computationally intensive.
Meanwhile, their controllability is mostly limited to single-composer conditioning. 
As a result, current text-based controls \citep{DBLP:journals/corr/abs-2409-11753} cannot generalize to more flexible scenarios, where, for example, generating music that blends stylistic traits of Bach and Chopin, or explicitly excluding the influence of Mozart.

Recent studies show that high-level concepts and behaviors admit differentiable structure in latent spaces.
Symbolic music score embeddings support multi-class and multi-label classification \citep{10.1145/3746278.3759392}.
SEAL finds that reasoning behavior is separable in LLM latent space \citep{chen2025sealsteerablereasoningcalibration}.
Motivated by this evidence, we hypothesize that composer style is differentially represented in symbolic music models and we empirically validate composer structure in Appendix~\ref{sec:latent_space}.

Therefore, building upon the limitations of prior training-based approaches and our validated representational hypothesis, we introduce \textbf{Composer Vector}, a latent-space steering method to control the composer style for symbolic music generation. 
Instead of training, Composer Vector operates directly within the model’s internal representation space to capture a stylistic vector corresponding to individual composers.
By adding or interpolating these vectors to music language models at inference time, we can continuously modulate the generated music toward or away from specific composer styles, allowing not only style transfer but also style fusion and style suppression.
Overall, our contributions are: (i) We introduce Composer Vector, a latent-space steering method that provides fine-grained control over composer style in symbolic music generation. (ii) We show that this approach enables continuous control over composer-style intensity, allowing smooth and interpretable modulation along a steering. (iii) We show that Composer Vector can be linearly combined to achieve multi-style fusion, enabling controllable mixtures of multiple composer styles.



\section{Composer Vector} %
\subsection{Style-steering Symbolic
Music Generation}
\begin{figure}[htbp]
    \centering
    \includegraphics[width=0.85\linewidth]{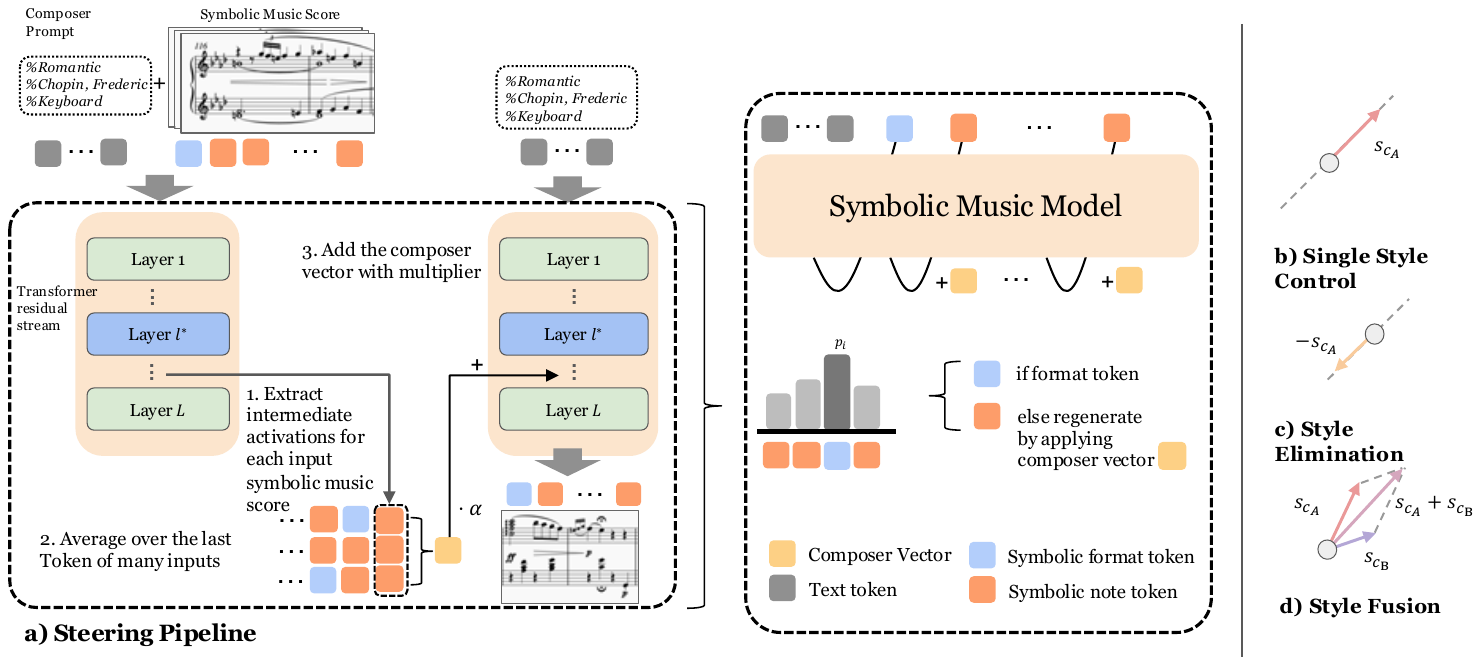}
  \caption{Inference-time control pipeline and style control diagrams}
  \label{fig:pipeline}
\end{figure}

We study composer style symbolic music generation with music language models such as NotaGen~\citep{wangNotaGenAdvancingMusicality2025b} and ChatMusician~\citep{yuanChatMusicianUnderstandingGenerating2024}.
Given a textual prompt $x$ (i.e., "\%Romantic \%Chopin, Frederic \%Keyboard" for NotaGen, or "Provide a musical piece that draws inspiration from Chopin's compositions." for ChatMusician), 
let $r\in\mathcal{C}$ denote the \textit{prompt composer} described in $x$, and let $c\in\mathcal{C}$ denote the \textbf{target composer}, where $\mathcal{C}$ is the set of composers.
Our proposed method allows controllable steering toward a  \emph{target} composer style $c$ at inference time, without additional training or fine-tuning, as illustrated in Figure~\ref{fig:pipeline}.

\subsection{Method}
\paragraph{Composer Vector Construction.} 
To obtain diverse stylistic contexts for each composer $c$, we first construct a prompt-piece corpus $\mathcal{D}_c = {\{x_i^c \oplus p_i^c\}}_{i=1}^{N_c}$ for each composer, where $x_i^c$ is a textual prompt describing the composer and $p_i^c$ is a symbolic score with ABC notation format\footnote{\url{https://abcnotation.com/}}.
The symbol $\oplus$ denotes concatenation of sequence.  
Both components contain composer style context, with the text providing explicit composer cues and the symbolic score encoding intrinsic musical style.
Given a transformer-based symbolic music generation model $f_\theta$, the hidden representation at layer $l$ for input $x_i^c \oplus p_i^c$ is denoted as $h^{(l)}(x_i^c \oplus p_i^c) \in \mathbb{R}^{T \times d}
$, where $T$ is the sequence length and $d$ is the hidden dimension.
Following a layer-wise analysis (Appendix~\ref{app:localization}), we identify an optimal layer $l^*$ that most strongly encodes composer-specific information, as determined by supervised linear probing and unsupervised clustering metrics.
\label{method:composer_vector}

At this layer, the \textbf{Composer Vector} for composer $c$ is defined as the mean embedding over their corpus:
$
    s_c \;=\; \frac{1}{N_c}\sum_{i=1}^{N_c} h^{(l^\star)}_T\!\big(x_i^c \oplus p_i^c\big)\in\mathbb{R}^{d},
$
This vector characterizes the latent direction corresponding to composer $c$.  
For any subset of composers $\mathcal{C}_k = \{c_1, \dots, c_k\}$, we compute a composite style vector by linear combination:
$
    s_{\mathcal{C}_k} = \sum_{i=1}^{k} w_{c_i}\, s_{c_i},
     w_{c_i} \in \mathbb{R}.
$
This enables smooth multi-style fusion and interpolation between composer styles.
Positive weights amplify the associated styles, whereas negative weights induce contrastive steering, suppressing certain stylistic influences.

\paragraph{Composer Style Steering.}
Given a target composer $c$ and prompt composer $r$, let $x$ as the text prompt corresponding to $r$.
We inject the corresponding Composer Vector $s_c$ into the model’s residual stream at layer $l^\star$ during generation. 
Let $\alpha \in \mathbb{R}$ denote the steering coefficient controlling the intensity of stylistic modulation. 
At each time step $t$, $y_{<t}$ denotes the output before $t$ of $f_\theta$. We modify the hidden state and rescale to preserve original norm:
$$
    \widehat{h}_t^{(l^\star)}(x,y_{<t}) = h_t^{(l^\star)}(x,y_{<t}) + \alpha \, s_c, 
\qquad
    \widehat{h}_t^{(l^\star)}(x,y_{<t}) \leftarrow 
        \frac{\|h_t^{(l^\star)}(x,y_{<t})\|_2}{\|\widehat{h}_t^{(l^\star)}(x,y_{<t})\|_2} 
        \, \widehat{h}_t^{(l^\star)}(x,y_{<t}).
$$
Finally, the model generates the output sequence under steering:
$
y = f_\theta(x; \, \alpha, s_c).
$
This norm-preserving steering allows continuous, directionally interpretable control over composer style.

\paragraph{Format Preservation.}
To maintain the structural integrity of symbolic notation, we incorporate a simple format preservation rule that limits steering to musically relevant tokens.  
At each decoding step $t$, if the predicted token belongs to a predefined set of format tokens (e.g., bar lines, measure separators, or control symbols), we preserve the model’s hidden state:
$
\widehat{h}_t^{(l^\star)}(x,y_{<t}) = h_t^{(l^\star)}(x,y_{<t}).
$
Otherwise, when the token corresponds to musical content (e.g., notes, durations, dynamics), we apply the composer vector steering as described above. 
This mechanism ensures that Composer Vector only affects expressive music content while preserving the score’s structural validity.

\section{Results}
We evaluate Composer Vector on NotaGen ~\citep{wangNotaGenAdvancingMusicality2025b} and ChatMusician \citep{yuanChatMusicianUnderstandingGenerating2024} on CLAP ~\citep{DBLP:conf/icassp/WuCZHBD23} and CLaMP \citep{wu2025clamp3universalmusic} similarity. Our experiments evaluate the effectiveness of the proposed \textbf{Composer Vector} for inference-time control of composer style in symbolic music generation. 
We assess three key aspects: (1) the effectiveness of style steering, (2) the continuity of control strength, and (3) the ability to perform multi-style fusion. Experiments setup details are in Appendix~\ref{app:exp_setup}

\paragraph{Controlling Symbolic Music Generation with Composer Vector.}

\begin{figure}[htbp]
    \centering    \includegraphics[width=0.8\linewidth]{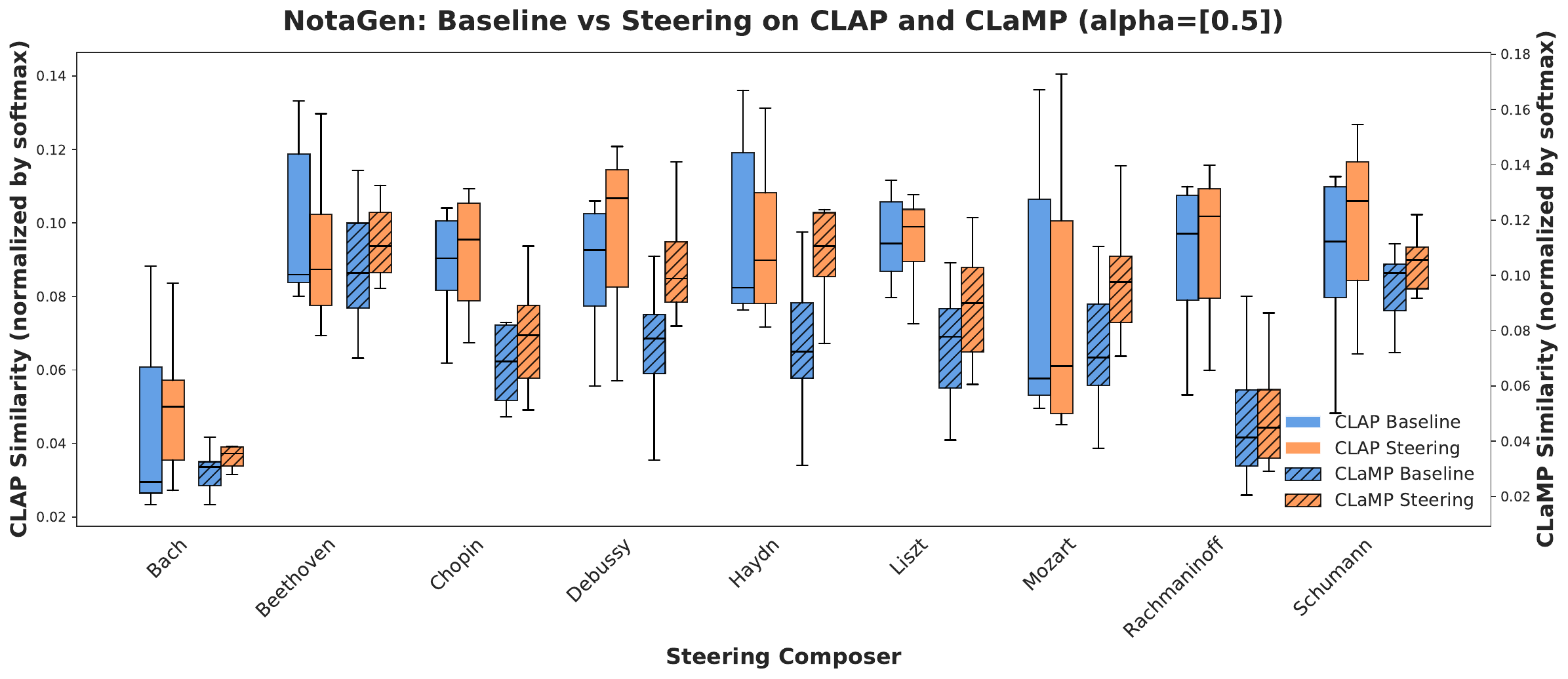}
    \caption{NotaGen: CLAP and CLaMP similarity before and after steering.}
    \label{fig:notagen_clap_clamp_a}
\end{figure}
In three objective evaluation metrics over two symbolic music generation models, we validate that the composer vector can guide composer style generation for symbolic music.
In similarity-based evaluation, we measure the similarity between generated scores and the target composer style before and after applying the Composer Vector.
For both models, similarity to the target composer style is consistently greater after steering than under the no-steering baseline, as shown in Figure~\ref{fig:notagen_clap_clamp_a} and Figure~\ref{fig:chat_clap_clamp_b}. 
This demonstrates that the Composer Vector shifts generation toward the target composer style.
This result confirms that composer vector approach enables controllable composer-style generation across different symbolic generation models. 
Beyond similarity metrics, we evaluate style control with a supervised composer classifier for music scores.
In Table~\ref{tab:classifier}, applying Composer Vector consistently increases the classifier’s predicted probability for the corresponding composer.
Moreover, the vector’s control exceeds the effect of the prompt: in ChatMusician, when the \textbf{target composer} is \textbf{Bach}, several prompt composers, such as Beethoven, Schumann, and Chopin, yield prediction probabilities exceeding 50\%.
This indicates that the Composer Vector encodes composer-specific information and dominates the prompt conditioning.
In conclusion, with similarity and classification metrics, we demonstrate Composer Vector can achieve fine-grained control across different composer styles.
\begin{table*}[htbp]
\centering
\small
\caption{Prediction probabilities (\%) of \textbf{target composer}, where the target composer is the one specified by the Composer Vector.
Columns are ordered by prompt era (Classical to Romantic).
Headers show the \textit{prompt} composer $r$ (italic) and the \textbf{target} composer $c$ (bold).
Each value indicates the classifier’s predicted probability that the output matches the target composer.}
\setlength{\tabcolsep}{5pt}
\renewcommand{\arraystretch}{1.15}
\scalebox{0.85}{
\begin{tabular}{lccccccccc}
\toprule
\textbf{Method} &
\shortstack{\textit{Bach}\\\textbf{Mozart}} &
\shortstack{\textit{Haydn}\\\textbf{Beethoven}} &
\shortstack{\textit{Haydn}\\\textbf{Liszt}} &
\shortstack{\textit{Beethoven}\\\textbf{Mozart}} &
\shortstack{\textit{Beethoven}\\\textbf{Bach}} &
\shortstack{\textit{Schubert}\\\textbf{Ravel}} &
\shortstack{\textit{Schumann}\\\textbf{Bach}} &
\shortstack{\textit{Chopin}\\\textbf{Bach}} &
\shortstack{\textit{Debussy}\\\textbf{Beethoven}} \\
\midrule
NotaGen                 & 5.9 & 26.3 & 18.7 & 4.9 & 11.7 & 6.6 & 11.2 & 14.9 & 33.7 \\
+vector          & \textbf{14.0} & \textbf{34.2} & \textbf{26.2} & \textbf{16.0} & \textbf{17.9} & \textbf{14.4} & \textbf{29.2} & \textbf{22.5} & \textbf{52.2} \\
ChatMusician            & 3.4 & 4.5 & 5.0 & 6.0 & 25.9 & 1.2 & 29.5 & 23.0 & 1.4 \\
+vector     & \textbf{16.4} & \textbf{30.1} & \textbf{15.1} & \textbf{13.2} & \textbf{54.3} & \textbf{14.6} & \textbf{57.0} & \textbf{53.4} & \textbf{20.8} \\
\bottomrule
\end{tabular}
}
\label{tab:classifier}
\end{table*}




\paragraph{Composer Vector can Provide Continuous Style Control.}

We examine how the classifier probability of the target composer varies with the coefficient $\alpha$ in Figure~\ref{fig:steering_correlation}.
With Beethoven, Chopin, and Rachmaninoff as targets, we apply their vectors across multiple prompts and plot one regression line per prompt.
Across all prompts, $\alpha$ shows a consistent positive correlation with the target composer probability. 
Relative to the unsteered baseline (dashed lines), the curves lie higher for $\alpha>0$, showing a clear gain from steering.
For instance, in the Rachmaninoff-target case, the probability grows steadily and, by $\alpha{=}0.8$, is more than twice the value at $\alpha{=}0.1$.
Overall, $\alpha$ provides a controllable knob for style-intensity and larger values strengthen target-style features.

\paragraph{Multi-style Fusion through Composer Vector.}

We study style fusion by linearly combining two composer vectors and measuring classifier probabilities across coefficient pairs in Figure~\ref{fig:notagen-fusion}.
For instance, in the Debussy–Mozart fusion pair, the two regression lines diverge strongly in opposite directions as the coefficient changes.
As coefficient of Debussy increases, the corresponding probability increases while the Mozart's decreases, indicating continuous and interpretable interpolation.
This opposite–slope pattern recurs across panels with different prompt and steering coefficient.
In summary, across all fusion pairs, one composer’s probability consistently increases while the other decreases, demonstrating that linear composition of composer vectors enables continuous, interpretable style interpolation.

\section{Conclusion}
In this work, we proposed \textbf{Composer Vector}, an inference-time latent-space method for composer-style generation.
It operates in the model’s latent space to steer, blend, and fine-grained control styles without fine-tuning.
Our analysis showed that composer identity is explicitly encoded in deep layers, providing reliable directions for control.
Across two symbolic generation models and multiple metrics, Composer Vector can shift outputs toward a target composer. It supports continuous strength control via a scalar coefficient, and enables linear fusion of multiple composer styles.
Overall, these results establish a simple and interpretable mechanism for flexible composer-style control in symbolic music generation.


\newpage

\bibliographystyle{plainnat}
\bibliography{references}

\newpage
\appendix

\section{Related work}
\label{app:related}

\paragraph{Symbolic music generation.}
Symbolic music generation has long been an active area of research \citep{leNaturalLanguageProcessing2025}. By representing compositions through abstract notations, symbolic music allows researchers and creators to directly manipulate musical elements, and it enables the analysis and modeling of high-level musical concepts such as harmony, structure, and style \citep{wangNotaGenAdvancingMusicality2025b}. To realize these advantages in generation, a variety of symbolic representation systems have been developed, including MIDI \citep{bhandari2024text2midigeneratingsymbolicmusic, le2025meteormelodyawaretexturecontrollablesymbolic}, REMI \citep{lu2023musecocogeneratingsymbolicmusic, huang2020popmusictransformerbeatbased, vonrütte2024figarogeneratingsymbolicmusic}, and ABC notation \citep{yuanChatMusicianUnderstandingGenerating2024, wuCLaMPContrastiveLanguageMusic2023, wuCLaMP2Multimodal, wu2025clamp3universalmusic, wangNotaGenAdvancingMusicality2025b}. Although differing in granularity and design choices, these representations have each proven effective for symbolic music generation. For example, MIDI-based models can capture detailed performance signals, REMI emphasizes sequential regularity, and ABC notation has shown strong results in modeling melody and structure. Together, these representations make symbolic generation not only suitable for producing melodies but also capable of modeling expressive performance signals, thereby offering both controllability and expressiveness in music generation.

\paragraph{Inference time intervention.}
With the increasing prevalence of large language models  \citep{kaplan2020scalinglawsneurallanguage, yuan2025nativesparseattentionhardwarealigned} and the growing strength of base models, a line of research has investigated training-free inference-time steering as a way to adapt model behavior without additional fine-tuning  (\citep{wuImprovedRepresentationSteering2025, xu2025easyedit2easytousesteeringframework, rimsky-etal-2024-steering}). Specifically, inference-time steering methods can be broadly categorized into three families. Prompt-based approaches guide generation by designing or augmenting input prompts \citep{wuImprovedRepresentationSteering2025}. Activation-based interventions modify internal activations at inference time to steer latent representations toward desired directions \citep{Zou2023RepresentationEA, Bartoszcze2025RepresentationEF}. Decoding-based techniques instead adjust the sampling or decoding process to bias outputs toward specific attributes \citep{liang2024controllabletextgenerationlarge}. Moreover, these steering methods can also support the combination of multiple features, thereby enabling more efficient and fine-grained control of model outputs \citep{xu2025easyedit2easytousesteeringframework}.

\paragraph{Mechanistic interpretability.}
Mechanistic interpretability (MI) seeks to reverse engineer model computations into human-understandable algorithms by decomposing large models into smaller components and elementary operations \citep{raiPracticalReviewMechanistic2025}. Following \citep{Olah2020ZoomIA}, MI research is commonly organized around three fundamental objects of study. First, features are human-interpretable properties encoded in model activations (e.g., the token “dog” activating concepts such as “animal” or “pet”), and MI aims to decode these representations. Second, circuits are computational pathways that connect features or transformer components to implement specific behaviors, ranging from toy examples like the induction circuit \citep{elhage2021mathematical} to generalized notions of information flow. Third, universality investigates whether identified features or circuits recur across different models and tasks, which has critical implications for transferring insights from small-scale studies to large language models \citep{Olah2020ZoomIA, raiPracticalReviewMechanistic2025}.

\section{Composer Styles are Distinguishable in the Latent Space}
\label{sec:latent_space}
\begin{figure}[htbp]
    \centering
    \includegraphics[width=\linewidth]{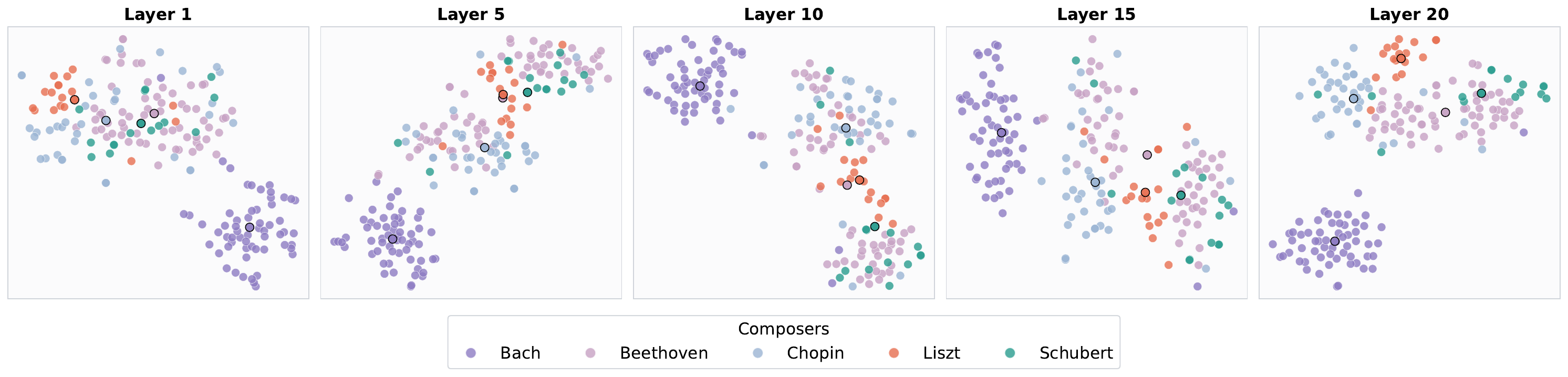}
    \caption{t-SNE visualization of piece-level embeddings across composers.}
    \label{fig:embeddings}
\end{figure}

We hypothesize that composer style, as a high-level musical attribute, is encoded in the latent representations of symbolic music models in a differentiable and structured manner.
Therefore, before introducing our steering method, we first examine whether composer styles are inherently encoded within the model’s latent representations. 
Understanding this structure is crucial, as effective steering relies on the existence of disentangled, style-specific directions in the representation space. 
Following a similar approach to \citet{10.1145/3746278.3759392}, we analyze how composer information is distributed across the layers of NotaGen. For each music piece $p$ in ABC notation \footnote{\url{https://abcnotation.com/}}, we extract its piece-level embedding defined as the hidden state of the final token at position $T$ for each layer $l$, denoted as $h^{(l)}_{T}(p)$, for each music piece $p$. We then visualize these embeddings using t-SNE~\citep{maaten2008visualizing}, as shown in Figure~\ref{fig:embeddings}.
Several key observations emerge: 
(i) Distinct clusters corresponding to different composers are clearly visible in the latent space, indicating that stylistic cues are represented internally. 
(ii) The separation is weak in lower layers but becomes progressively sharper in deeper ones, with the final layer exhibiting the most pronounced clustering. This aligns with prior findings that early layers capture local musical attributes (e.g., pitch transitions), whereas deeper layers encode higher-level structural and stylistic features~\citep{jawahar2019does}.  
(iii) Certain composers, such as Bach and Liszt, form more compact and well-separated clusters than others, reflecting their stronger stylistic distinctiveness. 
Overall, these results suggest that composer identity is explicitly manifested in the model’s hidden representations—providing a strong foundation for constructing and manipulating Composer Vector in subsequent sections.

\section{Experiments Details}
We evaluate the proposed \textbf{Composer Vector} method using two symbolic music generation models that operate in ABC notation: \textbf{NotaGen}~\citep{wangNotaGenAdvancingMusicality2025b} and \textbf{ChatMusician}~\citep{yuanChatMusicianUnderstandingGenerating2024}.
The evaluation dataset is derived from the \textbf{ASAP dataset}~\citep{asap-dataset}, which contains symbolic music annotated with composer identities. 
To ensure consistency across models, we convert all MusicXML files in the dataset into ABC format.

We focus on 11 composers spanning the Baroque, Classical, and Romantic periods.
For each composer, we generate symbolic pieces across three instrumental categories: \emph{keyboard}, \emph{chamber}, and \emph{orchestral}. 
Composer Vector are injected at inference time with steering coefficients $\alpha \in [0, 1]$. 
This setup enables systematic investigation of both single-composer steering and multi-composer fusion under different control intensities.

\noindent The experiments aim to address three research questions (RQs):
\begin{itemize}
    \item \textbf{RQ1:} How effectively can the Composer Vector control the generated composer style?
    \item \textbf{RQ2:} Can the Composer Vector provide \emph{continuous} and interpretable style control?
    \item \textbf{RQ3:} Can the Composer Vector achieve \emph{multi-style fusion} by combining stylistic directions?
\end{itemize}

\subsection{Experiment Design}
To answer the above questions, we design three groups of experiments corresponding to RQ1–RQ3.

\paragraph{How effective is the Composer Vector?}  
We assess whether the Composer Vector can guide generation toward the target composer style. 
For each prompt corresponding to a composer $c_p$, we apply Composer Vector from all 11 composers $\{c_1, \dots, c_{11}\}$ during generation. 
Each model (NotaGen and ChatMusician) produces 11 steered generations per prompt under steering coefficients $\alpha \in \{0.1, 0.3, 0.5, 0.8\}$. 
Composer Vector are extracted by averaging hidden representations of real ABC samples from the corresponding composer (see Section~\ref{method:composer_vector}). 
After obtaining the steering vectors, we apply them to the generation process for 11 composers, using steering coefficients of 0.1, 0.3, 0.5, and 0.8.

\paragraph{Can the Composer Vector provide continuous style control?}  
To verify continuous steering effects, we vary the coefficient continuously from 0.0 to 1.0 in 0.05 steps.  
We perform this test on five composers: Beethoven, Chopin, and Rachmaninoff.

\paragraph{Can the Composer Vector provide fusion of different styles?}  
We test combinations of composers to evaluate fusion effects.  
Mozart, Debussy, and Rachmaninoff are chosen as steering composers due to their distinct styles.  
We experiment with different fusion ratios: 0.1, 0.3, 0.5, 0.7, and 0.9.

\subsection{Evaluation}

\paragraph{Similarity-based Evaluation.}

We use CLAP \citep{DBLP:conf/icassp/WuCZHBD23} and CLaMP \citep{wu2025clamp3universalmusic} scores.  
For CLAP, we convert generated ABC scores into audio and embed the clips using the CLAP model.  
For CLaMP, we use the CLAMP3 model to embed symbolic scores directly. 
After obtaining embeddings from both audio and symbolic domains, we calculate the cosine similarity between the steered and original composer-style generations.  
An increase in similarity indicates that the steered generation is closer to the target composer's style.

\paragraph{Classification-based Evaluation.}
We also train a linear classifier to measure the style similarity more precisely. We use CLaMP3 \citep{wu2025clamp3universalmusic} as symbolic music encoder. We train the classifier on MIDI text format on 11 composers. We train the model on a large open source MIDI dataset \footnote{\url{https://huggingface.co/datasets/drengskapur/midi-classical-music}}. We separate the dataset into 70\%, 10\%, 20\%, training, validation, and testing set. The test accuracy is 89.38\% over 11 catagories.

\label{app:exp_setup}
\begin{table}[htbp]
\centering
\begin{threeparttable}
\caption{ASAP dataset summary: 1{,}067 performances and 236 distinct scores spanning 15 Western classical piano composers.}
\label{tab:asap-summary}
\begin{tabular}{lccc}
\toprule
\textbf{Composer} & \textbf{MIDI Perf.} & \textbf{Audio Perf.} & \textbf{MIDI/XML Scores} \\
\midrule
Bach          & 169 & 152 & 59 \\
Balakirev     & 10  & 3   & 1  \\
Beethoven     & 271 & 120 & 57 \\
Brahms        & 1   & 0   & 1  \\
Chopin        & 289 & 108 & 34 \\
Debussy       & 3   & 3   & 2  \\
Glinka        & 2   & 2   & 1  \\
Haydn         & 44  & 16  & 11 \\
Liszt         & 121 & 48  & 16 \\
Mozart        & 16  & 5   & 6  \\
Prokofiev     & 8   & 0   & 1  \\
Rachmaninoff  & 8   & 4   & 4  \\
Ravel         & 22  & 0   & 4  \\
Schubert      & 62  & 44  & 13 \\
Schumann      & 28  & 7   & 10 \\
Scriabin      & 13  & 7   & 2  \\
\midrule
\textbf{Total} & \textbf{1067} & \textbf{519} & \textbf{222} \\
\bottomrule
\end{tabular}
\end{threeparttable}
\end{table}

\section{Result Details}
\subsection{Composer Style Localization}
\label{app:localization}
We extract layer-wise hidden states from NotaGen and evaluate their ability to capture composer-specific style information. 
To this end, we employ a comprehensive evaluation framework that combines both supervised and unsupervised metrics to assess how well different layers organize musical pieces by composer identity.

\paragraph{Supervised evaluation.}  
We employ \textbf{linear probe accuracy}, a widely adopted method in mechanism interpretability~\citep{raiPracticalReviewMechanistic2025}, which quantifies how well a simple classifier can recover composer identity from layer-wise embeddings. 
Specifically, we train a logistic regression classifier with a 75\%/25\% train–test split (stratified by composer), using the test set accuracy as the evaluation score. 
This provides a direct measure of the discriminative power of each layer’s representation.

\paragraph{Unsupervised evaluation.}  
To complement the supervised probe, we compute six clustering-based metrics that capture different aspects of organization:    
(i) \textbf{k-nearest-neighbor purity (kNN-P, $k{=}5$)}, which evaluates local neighborhood consistency by checking the proportion of nearest neighbors with the same composer label
(ii) the \textbf{Davies–Bouldin index (DBI)}\footnote{\url{https://en.wikipedia.org/wiki/Davies-Bouldin_index}}, reported in negated form ($-\mathrm{DBI}$) so that larger values indicate better cluster separation; 
(iii) the \textbf{separation ratio (SepR)}, defined as the ratio of inter-cluster to intra-cluster distances, providing a normalized measure of relative separation; 

Together, these supervised and unsupervised metrics provide a comprehensive assessment of how well different layers encode composer-specific style.

\begin{figure}[t]
    \centering
    \begin{subfigure}[t]{0.9\linewidth}
        \centering
        \includegraphics[width=\linewidth]{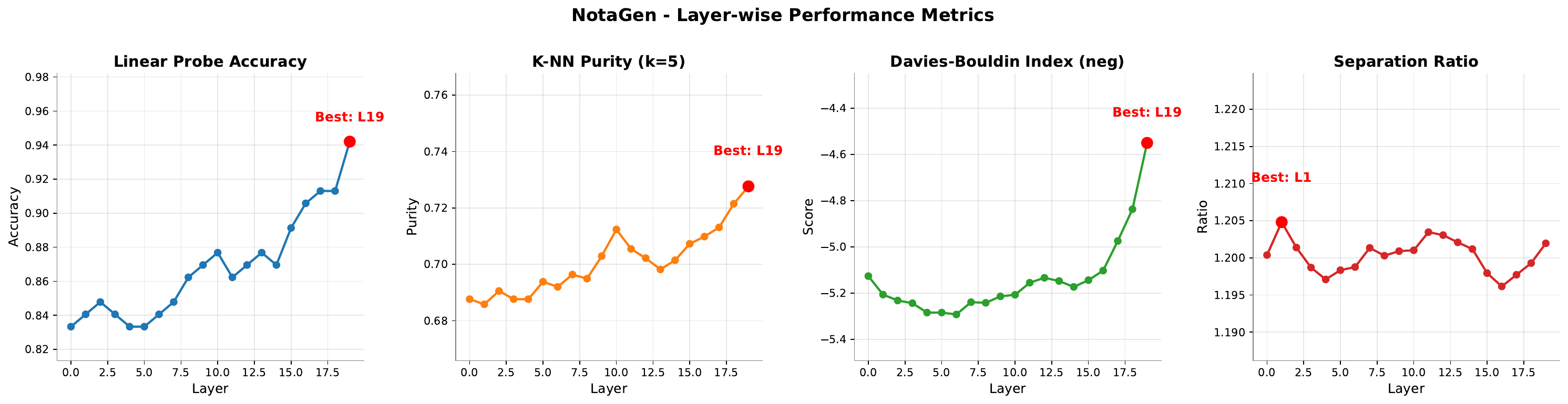}
        \caption{Layer-wise evaluation of composer style embeddings for NotaGen}
        \label{fig:notagen_extract}
    \end{subfigure}
    \vspace{0.6em}

    \begin{subfigure}[t]{0.9\linewidth}
        \centering
        \includegraphics[width=\linewidth]{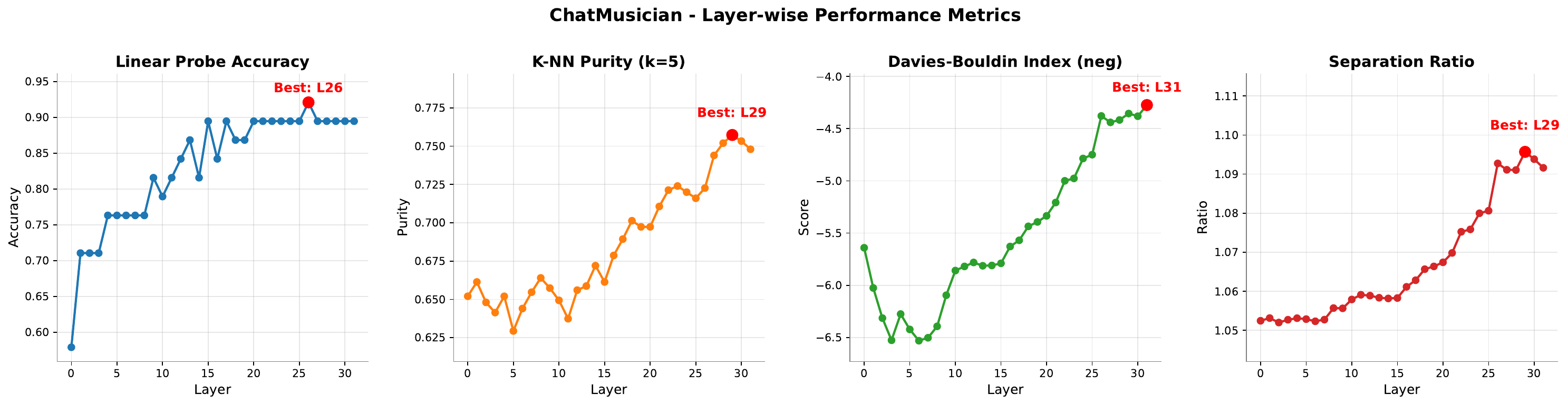}
        \caption{Layer-wise evaluation of composer style embeddings for ChatMusician}
        \label{fig:chatmusician_extract}
    \end{subfigure}

    \caption{Comparison of layer-wise composer style localization across different models.}
    \label{fig:layerwise_comparison}
\end{figure}

We select the layer that achieves the highest number of first-place rankings across the evaluation metrics as the final steering layer. In Figure~\ref{fig:notagen_extract}, the linear probe accuracy reaches to 94\% at layer 19th, which is the last layer of the NotaGen patch model. As for the ChatMusician, it has the highest K-NN purity (k=5) at layer 29th. These layers are the deeper layers in the model, which is consistent with the previous omposer style analysis. Overall, we choose 19th layer as steering layer in the NotaGen, and 29th layer in the ChatMusician as steering layer.

\subsection{Single-style steering}
\paragraph{Similarity-based Evaluation.}
\begin{figure}[htbp]
    \centering
    \includegraphics[width=0.85\linewidth]{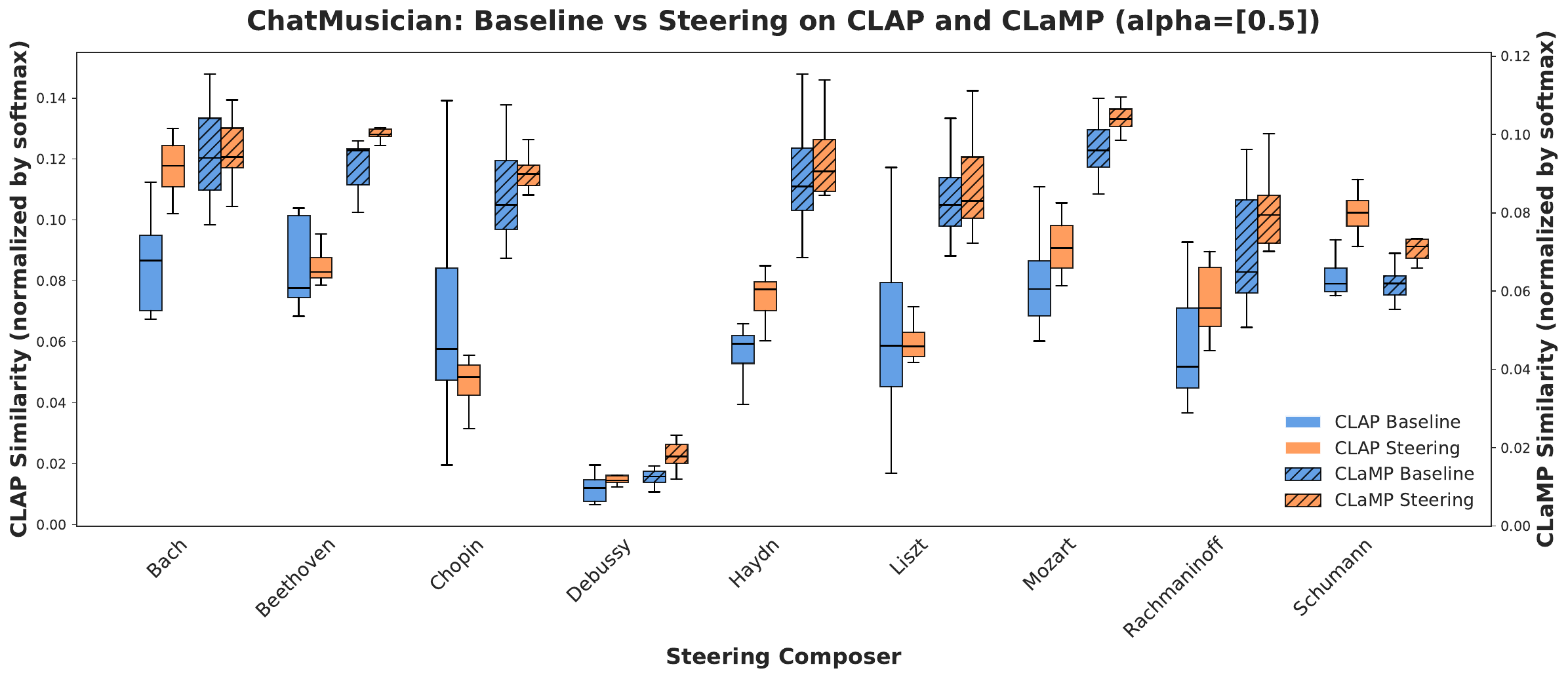}
    \caption{ChatMusician: CLAP and CLaMP similarity before and after steering.}
    \label{fig:chat_clap_clamp_b}
\end{figure}
\paragraph{Classification results.}

\begin{figure}[htbp]
    \centering
    \begin{subfigure}{0.48\linewidth}
        \centering
        \includegraphics[width=\linewidth]{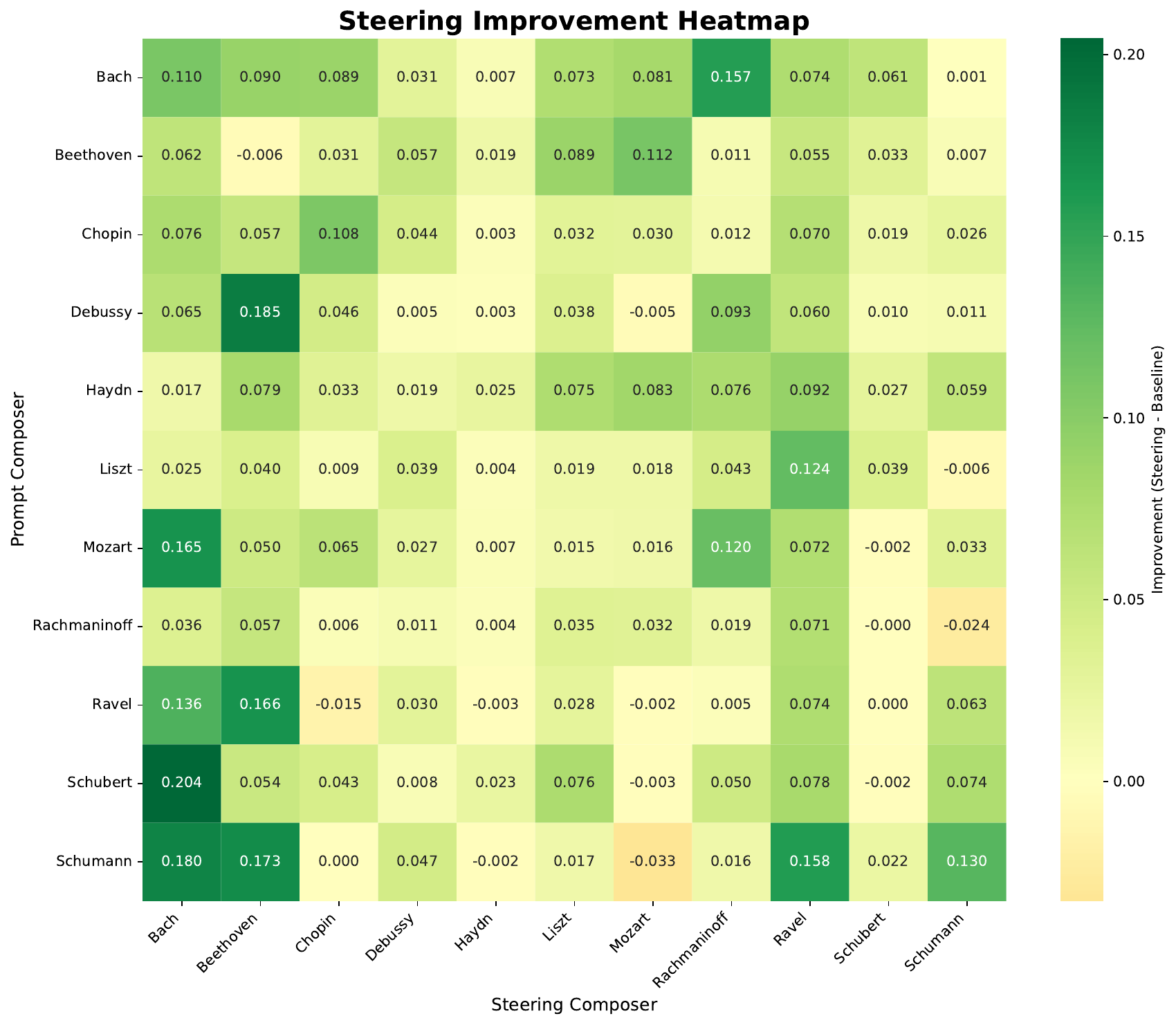}
        \caption{NotaGen.}
        \label{fig:notagen_heatmap}
    \end{subfigure}
    \hfill
    \begin{subfigure}{0.48\linewidth}
        \centering
        \includegraphics[width=\linewidth]{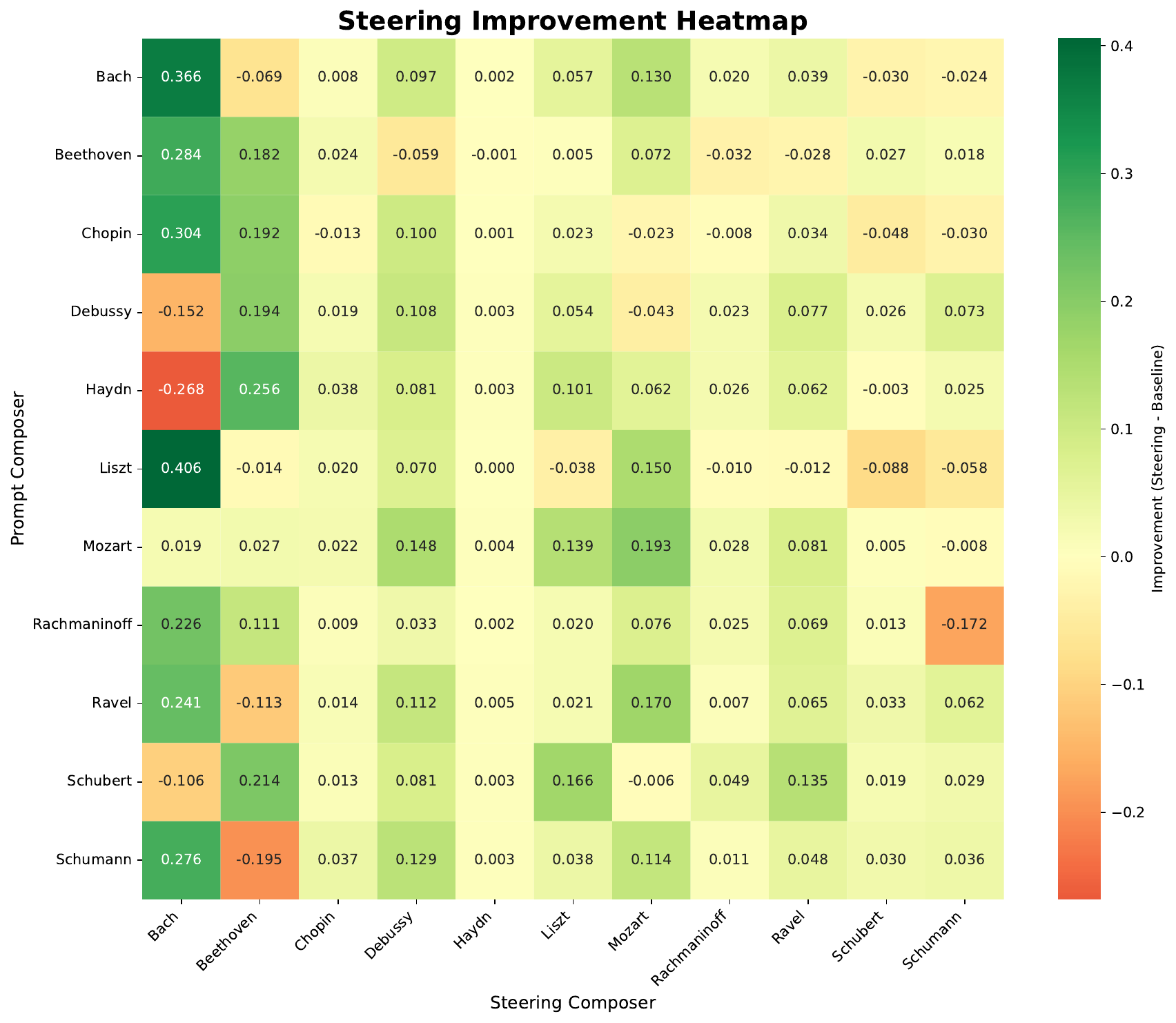}
        \caption{ChatMusician.}
        \label{fig:chat_heatmap}
    \end{subfigure}
    \caption{Steering improvement heatmaps for (a) NotaGen and (b) ChatMusician.}
    \label{fig:steering_improvement_heatmap}
\end{figure}

\begin{table*}[htbp]
\centering
\small
\setlength{\tabcolsep}{5pt}
\renewcommand{\arraystretch}{1.2}
\scalebox{0.8}{
\begin{tabular}{l l l l l l l l l l l l}
\toprule
\textbf{Prompt (B / S)} & \textbf{Bach} & \textbf{Beethoven} & \textbf{Chopin} & \textbf{Debussy} & 
\textbf{Haydn} & \textbf{Liszt} & \textbf{Mozart} & \textbf{Rach.} & 
\textbf{Ravel} & \textbf{Schubert} & \textbf{Schumann} \\
\midrule
Bach & 48.4\% / & 4.3\% / & 3.1\% / & 7.6\% / & 1.6\% / & 9.3\% / & 5.9\% / & 2.5\% / & 2.6\% / & 11.8\% / & 3.0\% / \\
     & \colorbox{Mycolor-green}{59.3\%} & \colorbox{Mycolor-green}{13.3\%} & \colorbox{Mycolor-green}{12.0\%} & \colorbox{Mycolor-green}{10.7\%} & 2.3\% & \colorbox{Mycolor-green}{16.5\%} & \colorbox{Mycolor-green}{14.0\%} & \colorbox{Mycolor-green}{18.2\%} & 10.0\% & \colorbox{Mycolor-green}{17.9\%} & 3.1\% \\
Beethoven & 11.7\% / & 35.8\% / & 6.4\% / & 4.5\% / & 1.2\% / & 12.3\% / & 4.9\% / & 7.2\% / & 4.7\% / & 4.3\% / & 7.0\% / \\
     & \colorbox{Mycolor-green}{17.9\%} & \colorbox{Mycolor-green}{35.3\%} & 9.5\% & \colorbox{Mycolor-green}{10.3\%} & 3.1\% & \colorbox{Mycolor-green}{21.2\%} & \colorbox{Mycolor-green}{16.0\%} & 8.3\% & \colorbox{Mycolor-green}{10.2\%} & 7.6\% & 7.7\% \\
Chopin & 14.9\% / & 28.6\% / & 12.8\% / & 4.9\% / & 0.6\% / & 9.9\% / & 4.0\% / & 10.2\% / & 3.3\% / & 1.1\% / & 9.6\% / \\
     & \colorbox{Mycolor-green}{22.5\%} & \colorbox{Mycolor-green}{34.3\%} & \colorbox{Mycolor-green}{23.6\%} & 9.3\% & 0.9\% & \colorbox{Mycolor-green}{13.0\%} & 6.9\% & \colorbox{Mycolor-green}{11.4\%} & \colorbox{Mycolor-green}{10.3\%} & 3.1\% & \colorbox{Mycolor-green}{12.2\%} \\
Debussy & 18.5\% / & 33.7\% / & 6.2\% / & 7.5\% / & 0.8\% / & 6.1\% / & 7.2\% / & 6.9\% / & 7.6\% / & 0.6\% / & 4.9\% / \\
     & \colorbox{Mycolor-green}{25.1\%} & \colorbox{Mycolor-green}{52.2\%} & \colorbox{Mycolor-green}{10.9\%} & 7.9\% & 1.0\% & 9.9\% & 6.7\% & \colorbox{Mycolor-green}{16.2\%} & \colorbox{Mycolor-green}{13.6\%} & 1.5\% & 6.0\% \\
Haydn & 16.8\% / & 26.3\% / & 4.3\% / & 5.2\% / & 1.6\% / & 18.7\% / & 10.7\% / & 3.4\% / & 4.7\% / & 4.3\% / & 4.1\% / \\
     & \colorbox{Mycolor-green}{18.5\%} & \colorbox{Mycolor-green}{34.2\%} & 7.6\% & 7.1\% & 4.0\% & \colorbox{Mycolor-green}{26.2\%} & \colorbox{Mycolor-green}{19.0\%} & \colorbox{Mycolor-green}{11.0\%} & \colorbox{Mycolor-green}{13.8\%} & 7.1\% & 10.0\% \\
Liszt & 19.6\% / & 26.9\% / & 11.0\% / & 4.1\% / & 1.2\% / & 11.0\% / & 2.9\% / & 7.1\% / & 6.6\% / & 0.9\% / & 8.8\% / \\
     & \colorbox{Mycolor-green}{22.1\%} & \colorbox{Mycolor-green}{30.9\%} & \colorbox{Mycolor-green}{11.9\%} & 8.0\% & 1.6\% & \colorbox{Mycolor-green}{12.8\%} & 4.7\% & \colorbox{Mycolor-green}{11.4\%} & \colorbox{Mycolor-green}{19.0\%} & 4.8\% & 8.1\% \\
Mozart & 11.1\% / & 23.3\% / & 3.9\% / & 5.4\% / & 3.1\% / & 24.5\% / & 11.2\% / & 3.5\% / & 4.6\% / & 4.3\% / & 4.9\% / \\
     & \colorbox{Mycolor-green}{27.6\%} & \colorbox{Mycolor-green}{28.3\%} & \colorbox{Mycolor-green}{10.4\%} & 8.2\% & 3.8\% & \colorbox{Mycolor-green}{26.1\%} & \colorbox{Mycolor-green}{12.9\%} & \colorbox{Mycolor-green}{15.5\%} & \colorbox{Mycolor-green}{11.9\%} & 4.1\% & 8.2\% \\
Rach. & 16.0\% / & 27.5\% / & 13.8\% / & 5.1\% / & 0.7\% / & 7.6\% / & 1.7\% / & 9.5\% / & 5.3\% / & 2.0\% / & 10.8\% / \\
     & \colorbox{Mycolor-green}{19.6\%} & \colorbox{Mycolor-green}{33.2\%} & \colorbox{Mycolor-green}{14.4\%} & 6.2\% & 1.1\% & \colorbox{Mycolor-green}{11.1\%} & 5.0\% & \colorbox{Mycolor-green}{11.3\%} & \colorbox{Mycolor-green}{12.4\%} & 2.0\% & 8.4\% \\
Ravel & 16.2\% / & 22.1\% / & 9.7\% / & 7.5\% / & 1.2\% / & 7.3\% / & 10.5\% / & 10.3\% / & 7.3\% / & 1.5\% / & 6.4\% / \\
     & \colorbox{Mycolor-green}{29.8\%} & \colorbox{Mycolor-green}{38.7\%} & 8.2\% & \colorbox{Mycolor-green}{10.4\%} & 0.9\% & \colorbox{Mycolor-green}{10.1\%} & \colorbox{Mycolor-green}{10.3\%} & \colorbox{Mycolor-green}{10.8\%} & \colorbox{Mycolor-green}{14.7\%} & 1.5\% & \colorbox{Mycolor-green}{12.7\%} \\
Schubert & 14.4\% / & 30.0\% / & 7.9\% / & 5.8\% / & 0.9\% / & 7.8\% / & 5.8\% / & 9.6\% / & 6.6\% / & 2.9\% / & 8.2\% / \\
     & \colorbox{Mycolor-green}{34.9\%} & \colorbox{Mycolor-green}{35.4\%} & \colorbox{Mycolor-green}{12.2\%} & 6.7\% & 3.2\% & \colorbox{Mycolor-green}{15.3\%} & 5.5\% & \colorbox{Mycolor-green}{14.7\%} & \colorbox{Mycolor-green}{14.4\%} & 2.7\% & \colorbox{Mycolor-green}{15.6\%} \\
Schumann & 11.2\% / & 15.7\% / & 14.9\% / & 5.1\% / & 1.2\% / & 11.0\% / & 9.5\% / & 8.5\% / & 6.8\% / & 1.0\% / & 14.9\% / \\
     & \colorbox{Mycolor-green}{29.2\%} & \colorbox{Mycolor-green}{33.0\%} & \colorbox{Mycolor-green}{14.9\%} & 9.8\% & 1.0\% & \colorbox{Mycolor-green}{12.7\%} & 6.2\% & \colorbox{Mycolor-green}{10.2\%} & \colorbox{Mycolor-green}{22.7\%} & 3.2\% & \colorbox{Mycolor-green}{28.0\%} \\
\bottomrule
\end{tabular}}
\caption{All Classifier Prediction Results for NotaGen. Baseline (top, ends with “/”) and Steering (bottom). Numbers with green background indicate steering probabilities greater than 10\%. “Rach.” = Rachmaninoff.}
\label{tab:notagen-probability}
\end{table*}

\begin{table*}[htbp]
\centering
\small
\setlength{\tabcolsep}{5pt}
\renewcommand{\arraystretch}{1.2}
\scalebox{0.8}{
\begin{tabular}{l l l l l l l l l l l l}
\toprule
\textbf{Prompt (B / S)} & \textbf{Bach} & \textbf{Beethoven} & \textbf{Chopin} & \textbf{Debussy} & 
\textbf{Haydn} & \textbf{Liszt} & \textbf{Mozart} & \textbf{Rach.} & 
\textbf{Ravel} & \textbf{Schubert} & \textbf{Schumann} \\
\midrule
Bach & 15.1\% / & 31.9\% / & 0.8\% / & 9.1\% / & 0.1\% / & 12.6\% / & 3.4\% / & 4.4\% / & 9.7\% / & 6.1\% / & 6.7\% / \\
     & \colorbox{Mycolor-green}{51.7\%} & \colorbox{Mycolor-green}{25.0\%} & 1.6\% & \colorbox{Mycolor-green}{18.8\%} & 0.4\% & \colorbox{Mycolor-green}{18.4\%} & \colorbox{Mycolor-green}{16.4\%} & 6.5\% & \colorbox{Mycolor-green}{13.6\%} & 3.1\% & 4.3\% \\
Beethoven & 25.9\% / & 2.5\% / & 0.9\% / & 22.6\% / & 0.4\% / & 7.4\% / & 6.0\% / & 7.4\% / & 19.5\% / & 2.7\% / & 4.6\% / \\
     & \colorbox{Mycolor-green}{54.3\%} & \colorbox{Mycolor-green}{20.7\%} & 3.2\% & \colorbox{Mycolor-green}{16.7\%} & 0.4\% & 7.9\% & \colorbox{Mycolor-green}{13.2\%} & 4.2\% & \colorbox{Mycolor-green}{16.7\%} & 5.4\% & 6.4\% \\
Chopin & 23.0\% / & 6.2\% / & 3.8\% / & 9.9\% / & 0.4\% / & 10.4\% / & 15.3\% / & 5.1\% / & 10.3\% / & 7.2\% / & 8.5\% / \\
     & \colorbox{Mycolor-green}{53.4\%} & \colorbox{Mycolor-green}{25.4\%} & 2.5\% & \colorbox{Mycolor-green}{19.9\%} & 0.4\% & \colorbox{Mycolor-green}{12.7\%} & \colorbox{Mycolor-green}{13.0\%} & 4.3\% & \colorbox{Mycolor-green}{13.7\%} & 2.3\% & 5.5\% \\
Debussy & \colorbox{Mycolor-green}{63.8\%} / & 1.4\% / & 1.1\% / & 9.3\% / & 0.1\% / & 1.5\% / & 17.1\% / & 2.1\% / & 2.0\% / & 1.2\% / & 0.5\% / \\
     & \colorbox{Mycolor-green}{48.6\%} & \colorbox{Mycolor-green}{20.8\%} & 3.1\% & \colorbox{Mycolor-green}{20.1\%} & 0.4\% & 6.9\% & \colorbox{Mycolor-green}{12.8\%} & 4.3\% & 9.7\% & 3.8\% & 7.8\% \\
Haydn & \colorbox{Mycolor-green}{63.6\%} / & 4.5\% / & 1.5\% / & 5.7\% / & 0.1\% / & 5.0\% / & 3.7\% / & 1.8\% / & 8.5\% / & 3.1\% / & 2.8\% / \\
     & \colorbox{Mycolor-green}{36.7\%} & \colorbox{Mycolor-green}{30.1\%} & 5.3\% & \colorbox{Mycolor-green}{13.8\%} & 0.4\% & \colorbox{Mycolor-green}{15.1\%} & \colorbox{Mycolor-green}{9.9\%} & 4.4\% & \colorbox{Mycolor-green}{14.7\%} & 2.7\% & 5.3\% \\
Liszt & 20.8\% / & 15.5\% / & 0.5\% / & 10.4\% / & 0.2\% / & 9.2\% / & 2.8\% / & 5.8\% / & 12.8\% / & 12.4\% / & 9.7\% / \\
     & \colorbox{Mycolor-green}{61.3\%} & \colorbox{Mycolor-green}{14.1\%} & 2.5\% & \colorbox{Mycolor-green}{17.4\%} & 0.3\% & 5.4\% & \colorbox{Mycolor-green}{17.8\%} & 4.8\% & \colorbox{Mycolor-green}{11.7\%} & 3.6\% & 3.9\% \\
Mozart & \colorbox{Mycolor-green}{53.5\%} / & 22.3\% / & 0.9\% / & 3.0\% / & 0.1\% / & 4.5\% / & 2.7\% / & 3.2\% / & 3.1\% / & 1.2\% / & 5.6\% / \\
     & \colorbox{Mycolor-green}{55.4\%} & \colorbox{Mycolor-green}{25.0\%} & 3.1\% & \colorbox{Mycolor-green}{17.8\%} & 0.5\% & \colorbox{Mycolor-green}{18.4\%} & \colorbox{Mycolor-green}{21.9\%} & 6.1\% & \colorbox{Mycolor-green}{11.2\%} & 1.7\% & 4.8\% \\
Rach. & 29.4\% / & 7.3\% / & 1.1\% / & 15.8\% / & 0.2\% / & 7.2\% / & 6.7\% / & 3.2\% / & 4.5\% / & 2.6\% / & \colorbox{Mycolor-green}{22.0\%} / \\
     & \colorbox{Mycolor-green}{52.0\%} & \colorbox{Mycolor-green}{18.5\%} & 1.9\% & \colorbox{Mycolor-green}{19.1\%} & 0.3\% & 9.2\% & \colorbox{Mycolor-green}{14.3\%} & 5.8\% & \colorbox{Mycolor-green}{11.4\%} & 3.9\% & 4.8\% \\
Ravel & 40.4\% / & 24.0\% / & 0.9\% / & 6.4\% / & 0.1\% / & 3.6\% / & 8.0\% / & 4.1\% / & 4.2\% / & 1.6\% / & 6.8\% / \\
     & \colorbox{Mycolor-green}{64.5\%} & \colorbox{Mycolor-green}{12.6\%} & 2.3\% & \colorbox{Mycolor-green}{17.5\%} & 0.6\% & 5.7\% & \colorbox{Mycolor-green}{24.9\%} & 4.8\% & \colorbox{Mycolor-green}{10.7\%} & 4.9\% & \colorbox{Mycolor-green}{13.0\%} \\
Schubert & \colorbox{Mycolor-green}{71.0\%} / & 0.6\% / & 1.1\% / & 6.3\% / & 0.1\% / & 1.9\% / & 14.0\% / & 0.7\% / & 1.2\% / & 1.4\% / & 1.8\% / \\
     & \colorbox{Mycolor-green}{60.4\%} & \colorbox{Mycolor-green}{22.0\%} & 2.4\% & \colorbox{Mycolor-green}{14.3\%} & 0.4\% & \colorbox{Mycolor-green}{18.4\%} & \colorbox{Mycolor-green}{13.4\%} & 5.6\% & \colorbox{Mycolor-green}{14.6\%} & 3.4\% & 4.7\% \\
Schumann & 29.5\% / & \colorbox{Mycolor-green}{47.7\%} / & 0.7\% / & 3.3\% / & 0.2\% / & 2.8\% / & 5.4\% / & 3.5\% / & 3.8\% / & 0.6\% / & 2.5\% / \\
     & \colorbox{Mycolor-green}{57.0\%} & \colorbox{Mycolor-green}{28.1\%} & 4.5\% & \colorbox{Mycolor-green}{16.2\%} & 0.5\% & 6.6\% & \colorbox{Mycolor-green}{16.9\%} & 4.6\% & 8.6\% & 3.6\% & 6.1\% \\
\bottomrule
\end{tabular}}
\caption{All Classifier Prediction Results for ChaMusician. Baseline (top, ends with “/”) and Steering (bottom). Numbers with green background indicate steering probabilities greater than 10\%. “Rach.” = Rachmaninoff.}
\label{tab:chatmusician-probability}
\end{table*}

We also use the naive generation as baseline to compare the prediction probability difference. The Table~\ref{tab:notagen-probability} shows the probability difference for each generated score. 
For instance, a piece has prompt "Bach" and steered by "Mozart".
Then we will compare the Mozart probability versus the Mozart probability of naive "Bach" generation.
As shown in the Figure~\ref{fig:steering_improvement_heatmap}. 97.5\% of NotaGen probability difference is larger than 0. ChatMusician 84.3\% probability difference is larger than 0.
The composer vector is more stable on the NotaGen, but in general the steered generation will have higher probability for that steering composer.

\subsection{Steering Coefficient Analysis}

\begin{figure}
    \centering
    \includegraphics[width=\linewidth]{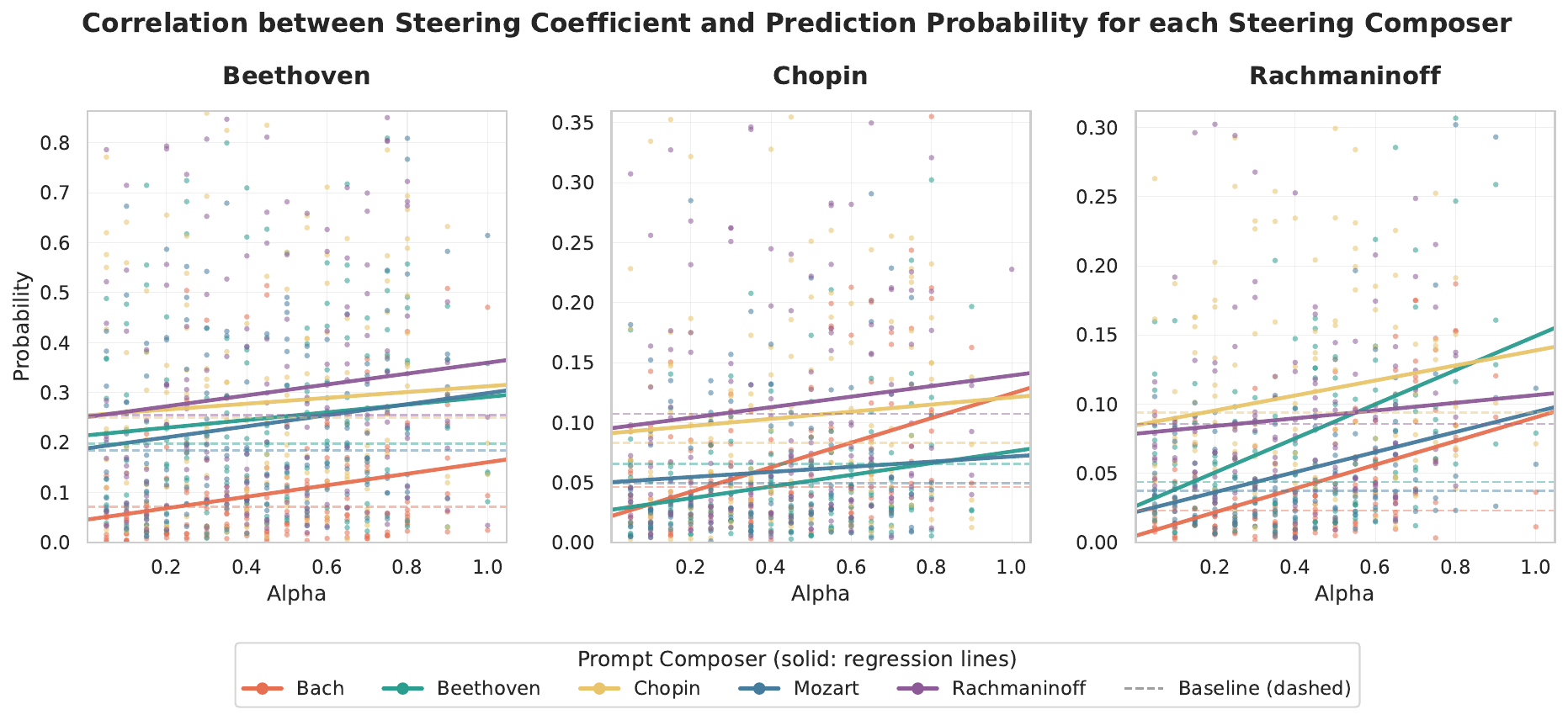}
    \caption{Correlation Between Steering Coefficient and Prediction Probability}
    \label{fig:steering_correlation}
\end{figure}
To further assess the effectiveness of the steering coefficient, we analyze how the classifier-predicted probability of the steering composer varies with the coefficient value $\alpha$.
Figure~\ref{fig:steering_correlation} illustrates the relationship between the steering coefficient and classifier-predicted probability for three representative composers: Beethoven, Chopin, and Rachmaninoff.
Each point represents a single steered sample, and the color indicates the prompt composer $r$ used during generation.
Each subfigure corresponds to one steering composer, where each point represents an individual steered sample, and the color indicates the prompt composer used during generation. 
Solid lines denote linear regression fits between $\alpha$ and the classifier-predicted probability, while the dashed line represents the unsteered baseline.

First, we observe a consistent positive correlation between the steering coefficient $\alpha$ and the classifier probability of the corresponding composer.
As $\alpha$ increases, the predicted likelihood of the target composer rises across nearly all prompt conditions. 
For instance, in the Rachmaninoff-steering group, the classifier probability grows steadily from near-baseline levels at $\alpha{=}0.1$ to more than double at $\alpha{=}0.8$ when base prompts are Bach, Mozart, and Beethoven, demonstrating that stronger steering coefficients effectively enhance stylistic alignment.
Second, cross-composer variations highlight meaningful stylistic distinctions among composers.
In the Beethoven-steering experiments, samples prompted by Bach yield notably lower classifier probabilities compared to those prompted by Romantic-era composers such as Liszt or Schumann.
This pattern suggests that stylistic distance between Bach and Beethoven constrains the model’s ability to transfer features through latent steering, whereas stylistically closer composers benefit more from steering amplification.

In summary, these results show that the steering coefficient $\alpha$ provides a controllable and interpretable mechanism for modulating composer-style intensity.
Larger coefficients reinforce stylistic features of the target composer, while inter-composer variations in response strength reveal underlying structure in the model’s latent style representation.

\subsection{Multi-style Fusion through Composer Vector}
To further assess the compositional flexibility of the proposed steering method, we investigate style fusion by linearly combining two composer vectors.
We focus on three stylistically distinct composers: Mozart, Debussy, and Rachmaninoff, whose contrasting harmonic and textural characteristics allow a clear examination of how linear blending influences stylistic interpolation.
This diversity allows us to isolate and contrast the steering effects among stylistically distant composers.

\begin{figure}
    \centering
    \includegraphics[width=0.95\linewidth]{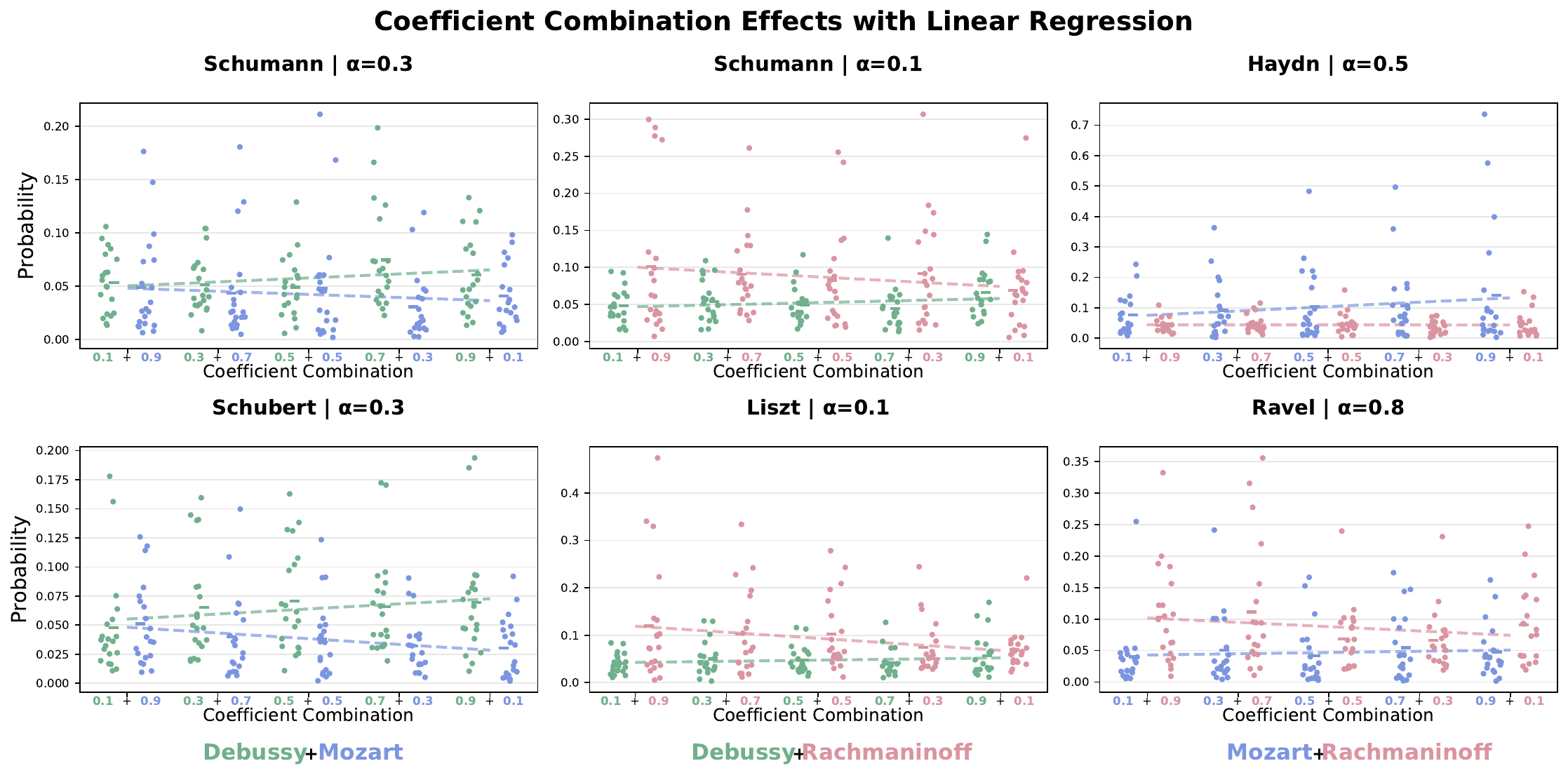}
    \caption{Probability of Steering Composers for Style Fusion}
    \label{fig:notagen-fusion}
\end{figure}

Figure~\ref{fig:notagen-fusion} shows the classifier-predicted probabilities for each steering composer across different coefficient combinations (0.1–0.9).
Each point represents an individual steered sample, and regression lines capture the trend between coefficient ratios and classifier probabilities.

First, in the Debussy–Mozart fusion pair, the two regression lines diverge strongly in opposite directions as the coefficient changes.
The probability of Debussy rises steadily while that of Mozart decreases, forming a clear linear correlation.
This pattern indicates that their latent style representations are highly separable, allowing precise control over stylistic balance between impressionistic fluidity and classical clarity.
Moreover, even in fusion pairs with weaker global trends, such as Debussy–Rachmaninoff and Mozart–Rachmaninoff in Figure~\ref{fig:notagen-fusion}, the highest classifier probabilities for individual samples still align with the dominant coefficient in each combination.
This pattern suggests that although the overall regression slopes are mild, the model retains localized sensitivity to the steering ratio, capturing the intended stylistic emphasis at the sample level.
In other words, the maximum probability values vary systematically with the coefficient ratio, indicating that the model’s responses remain well aligned with the intended style proportions.

In summary, across all pairs, one composer’s probability consistently increases while the other decreases, demonstrating that linear composition of composer vectors enables continuous, interpretable style interpolation.
The degree of slope separation reflects how linearly independent the underlying stylistic manifolds are in the model’s latent space, confirming that composer-vector fusion provides a controllable mechanism for generating hybrid musical styles.

\end{document}